 \documentclass[preprintnumbers,amsmath,amssymb,floatfix]{revtex4}

\usepackage{graphicx}
\usepackage{dcolumn}
\usepackage{bm}
\usepackage{amssymb}
\usepackage{epsfig}
\usepackage{color}

\newcommand{\ba}{\begin{eqnarray}}
\newcommand{\ea}{\end{eqnarray}}
\newcommand{\be}{\begin{equation}}
\newcommand{\ee}{\end{equation}}
\newcommand{\bdisplay}{\begin{displaymath}}
\newcommand{\edisplay}{\end{displaymath}}

\newcommand{\eq}[1]{Eq.\,(\ref{#1})}

\begin{document}

\title{Decoupling the NLO coupled DGLAP evolution equations: an analytic solution to pQCD}
\author{Martin~M.~Block}
\affiliation{Department of Physics and Astronomy, Northwestern University, 
Evanston, IL 60208}
\author{Loyal Durand}
\affiliation{Department of Physics, University of Wisconsin, Madison, WI 53706}
\author{Phuoc Ha}
\affiliation{Department of Physics, Astronomy and Geosciences, Towson University , Towson, MD 21252}
\author{Douglas W. McKay}
\affiliation{Department of Physics and Astronomy, University of Kansas, Lawrence, KS 66045} 
\date{\today}

\begin{abstract}
Using repeated Laplace transform techniques, along with newly-developed accurate numerical inverse Laplace transform algorithms \cite{inverseLaplace1,inverseLaplace2},   we transform the {\em coupled, integral-differential} NLO singlet DGLAP equations first into {\em coupled differential} equations, then into {\em coupled algebraic} equations, which we can solve iteratively.  After Laplace inverting the algebraic solution analytically, we  numerically invert the solutions of the decoupled differential equations. Finally, we arrive at the {\em decoupled} NLO evolved solutions 
\ba
F_s(x,Q^2)&=&{\cal F}_s(F_{s0}(x), G_0(x)),\nonumber\\
G(x,Q^2)&=&{\cal G}(F_{s0}(x), G_0(x)),\nonumber
\ea 
where ${\cal F}_s$ and $\cal G$ are known functions- determined using the DGLAP splitting functions up to NLO in the strong coupling constant $\alpha_s(Q^2)$. The functions $F_{s0}(x)\equiv F_s(x,Q_0^2)$ and $G_{0}(x)\equiv G(x,Q_0^2)$ are the starting functions for the evolution at $Q^2=Q_0^2$.  This approach furnishes us with a new tool for readily  obtaining, {\em independently},  the effects of the starting functions  on either the evolved gluon or singlet structure functions, as a function of both $Q^2$ and $Q_0^2$. It is not necessary to evolve coupled integral-differential equations numerically on a two-dimensional grid, as is currently done.  The same approach can be used for NLO  non-singlet distributions where it is simpler,  only requiring one Laplace transform.   We make successful NLO numerical comparisons to two non-singlet distributions, using NLO quark distributions published by the MSTW collaboration 
\cite{MSTW1}, over a large range of $x$ and $Q^2$.  Our method is readily generalized to higher orders in the strong coupling constant $\alpha_s(Q^2)$. 

\end{abstract}


\maketitle


\section{Introduction} \label{sec:introduction} 
In order to interpret the experimental results at the Large Hadron Collider---in the search for new physics---accurate knowledge of gluon distribution functions at small Bjorken $x$ and large virtuality $Q^2$  plays a vital role in estimating QCD backgrounds and in calculating gluon-initiated processes. 
The gluon and quark distribution functions have traditionally been determined  simultaneously by fitting experimental data on neutral- and charged-current deep inelastic scattering processes and some jet data over a large domain  of values of $x$ and $Q^2$.  The distributions at small $x$ and large $Q^2$ are determined mainly by the  proton structure function $F_2^{\gamma p}(x,Q^2)$ measured in deep inelastic $ep$ (or $\gamma^*p$) scattering.  The fitting process starts with an initial  $Q^2_0$, typically less than $m_c^2$, the square of the  $c$ quark mass of $\approx 2$ GeV$^2$, and  individual quark and gluon initial distributions which are parameterized with pre-determined shapes in $x$ determined by a set of  adjustable input  parameters---given as functions of $x$ for the chosen $Q_0^2$. The distributions are then evolved numerically on a two-dimensional grid in $x$ and $Q^2$  to larger $Q^2$ using the coupled integral-differential DGLAP equations  \cite{dglap1,dglap2,dglap3}, typically in leading order (LO) and next-to- leading order (NLO), and the results used to predict the measured quantities. The final distributions are then determined by adjusting the input parameters  to obtain a best fit to the data. This procedure is very indirect in the case of the gluon: the gluon distribution $G(x,Q^2) = xg(x,Q^2)$ does not contribute directly to the accurately determined structure function $F_2^{\gamma p}(x,Q^2)$, and is determined only through the quark distributions in conjunction with the evolution equations, or at large $x$, from jet data. For recent determinations of the gluon and quark distributions, see \cite{CTEQ6.1,CTEQ6.5,MRST,MRST4, MSTW1}.

In the following, we will summarize our analytic method that determines the singlet structure function $F_s(x,Q^2)$ and $G(x,Q^2)$ {\em directly } and {\em individually}, using as input  $F_{s0}(x)\equiv F_s(x,Q^2_0)$ and  $G_0(x)\equiv G(x,Q^2_0)$, where $Q_0^2$ is arbitrary, with the guarantee that each distribution individually satisfies the NLO coupled DGLAP equations.   

The method is extended to calculate NLO non-singlet functions, so that we can  also find individual quark {\em and} gluon distributions analytically in terms of the starting distributions  of the individual quark and gluon distributions.  We will also give some numerical examples for non-singlet NLO valence quark distributions, comparing them to the MSTW \cite{MSTW1} published NLO valence quark distributions. 


\section{NLO Singlet Sector}
Our approach uses an unusual  application of multiple Laplace transforms \cite{bdm1,bdm2}. In this note, we use {\em double} Laplace transforms,  first transforming the coupled DGLAP integral-differential equations into a set of coupled {\em differential} equations in Laplace space, and finally, into a set of coupled {\em algebraic} equations in a second Laplace space.  We then solve the coupled algebraic equations in this second Laplace space. To obtain our final results, we must {\em invert} the Laplace transforms. The second transform to the algebraic equation space is analytically invertible to the space in which we had the coupled differential equations. The final inversion, from this Laplace space back to our initial space,  must be obtained by numerical inverse Laplace transformations \cite{inverseLaplace1,inverseLaplace2}.   

We first introduce the variable $v\equiv \ln(1/x)$ into the  NLO coupled DGLAP equations. This turns them into coupled convolution equations in $v$ space, which, after introducing a new variable $\tau (Q^2,Q^2_0)={1\over 4\pi}\int_{Q_0^2}^{Q^2}\alpha_s(Q^2)d\,\ln Q^2$, are readily Laplace transformed to obtain a set of coupled {\em  homogeneous first-order differential}\, equations in the variable $\tau$. The parameters of these transformed  equations are known functions of $s$, the Laplace-space variable.  These equations are then Laplace transformed a {\em second } time, essentially transforming the  variable $\tau $ of the coupled differential equations  into a new Laplace variable $U$, with the resulting equations being coupled {\em algebraic} equations in $U$---with $s$ again being a parameter---which are then solved iteratively. These solutions, in $s$ and $U$,  are analytically Laplace inverted back to variables $s$ and $\tau$. Using fast and accurate numerical inverse Laplace transform algorithms \cite{inverseLaplace1,inverseLaplace2}, we transform the  solutions back into $v$ space, and, finally, into Bjorken $x$-space to obtain  
$F_s(x,Q^2)={\cal F}_s(F_{s0}(x), G_0(x))$ and
$G(x,Q^2)={\cal G}(F_{s0}(x), G_0(x))$, where the functions $\cal F$ and $\cal G$ are determined by the splitting functions in the DGLAP equations.

 A similar method was used in an earlier paper \cite{bdhm1} in which we obtained the decoupled solutions in LO for both the singlet and the non-singlet sector, using only one Laplace transform.  The $\tau$ dependence in that case was trivial, and the decoupled equations could be solved directly. The extra Laplace transform that appears in the present work is necessitated by the nontrivial  dependence of the NLO terms on $\tau$. 

Our method is readily generalized to all orders in the strong coupling constant, but for brevity we limit ourselves to NLO in this paper.  We write  the coupled NLO DGLAP equations  \cite{bdm1,bdm2} schematically, using the convolution symbol $\otimes$, as
\ba
\frac{4\pi}{\alpha_s(Q^2)}\frac{\partial F_s}{\partial\ln{Q^2}}(x,Q^2)&=& F_s\otimes \left( P_{qq}^0+\frac{\alpha_s(Q^2)}{4\pi}P_{qq}^1\right)(x,Q^2)+G\otimes \left( P_{qg}^0+\frac{\alpha_s(Q^2)}{4\pi}P_{qg}^1\right)(x,Q^2),\label{dFdtauofQsq}\\
\frac{4\pi}{\alpha_s(Q^2)}\frac{\partial G}{\partial \ln{Q^2}}(x,Q^2)&=& F_s\otimes \left( P_{gq}^0+\frac{\alpha_s(Q^2)}{4\pi}P_{gq}^1\right)(x,Q^2)+G\otimes \left( P_{gg}^0+\frac{\alpha_s(Q^2)}{4\pi}P_{qg}^1\right)(x,Q^2).\label{dGdtauofQsq}
\ea

The $P_{qq}^0(x),\ P_{qg}^0(x),\ P_{gq}^0(x)$ and $P_{gg}^0(x)$ used in \eq{dFdtauofQsq} and \eq{dGdtauofQsq} are the LO singlet splitting function and the  $P_{qq}^1(x),\ P_{qg}^1(x),\ P_{gq}^1(x)$ and $P_{gg}^1(x)$ are the NLO singlet splitting functions, with
$\alpha_s(Q^2)$ the NLO running strong coupling constant. 
It is standard procedure  to construct $\alpha_s(Q^2)$  assuming three massless quarks, $u,\ d$ and $s$, below the $c$-quark threshold,  adjusting the QCD parameter $\Lambda$ at each successive threshold in $Q^2$,  i,e, at $Q^2=M_c^2$ and $M_b^2$, so that $\alpha_s(Q^2)$ remains continuous when the number of quarks changes as heavy $c$ and $b$ quarks begin to contribute.

Introducing the variable changes 
\ba 
v&\equiv &\ln (1/x),\quad w\equiv\ln (1/z),\qquad 
\tau(Q^2,Q_0^2)\equiv {1\over 4\pi}\int_{Q^2_0}^{Q^2}\alpha_s(Q'^2)d\,\ln Q'^2,\quad \label{vofaandtau}
\ea
  and  the notation 
\ba
\hat F_s(v,\tau)&\equiv & F_s(e^{-v},Q^2), \qquad\hat G(v,\tau)\equiv G(e^{-v},Q^2),\label{FandGhat} 
\ea
we rewrite the above DGLAP equations in terms of the convolution integrals
\ba
\frac{\partial {\hat F}_s}{\partial \tau}(v,\tau)&=& \int_0^v {\hat F}_s(w,\tau) \left({\hat H}_{qq}(v-w) +\frac{\alpha_s(\tau)}{4\pi}{\hat H}_{qq}^1(v-w)\right)d\,w\nonumber\\
&&+ \int_0^v {\hat G}(w,\tau) \left({\hat H}_{qg}(v-w) +\frac{\alpha_s(\tau)}{4\pi}{\hat H}_{qg}^1(v-w)\right)d\,w,\label{dFdtauoftau}\\
\frac{\partial {\hat G}}{\partial \tau}(v,\tau)&=&\int_0^v {\hat F}_s(w,\tau) \left({\hat H}_{gq}(v-w) +\frac{\alpha_s(\tau)}{4\pi}{\hat H}_{gq}^1(v-w)\right)d\,w\nonumber\\
&&+ \int_0^v {\hat G}(w,\tau) \left({\hat H}_{gg}(v-w) +\frac{\alpha_s(\tau)}{4\pi}{\hat H}_{gg}^1(v-w)\right)d\,w,\label{dGdtauoftau}
\ea
 where 
\ba 
{\hat F}_s(v,\tau )&\equiv& F_s(e^{-v},\tau),\qquad \hat G(v,\tau )\equiv G(e^{-v},\tau),\label{FandGhat2}\\
{\hat H}_{qq}^0(v)&\equiv&e^{-v}P_{qq}^0(e^{-v}),\quad {\hat H}_{qg}^0(v)\equiv e^{-v}P_{qg}^0(e^{-v}),\quad {\hat H}_{gq}^0(v)\equiv e^{-v}P_{gq}^0(e^{-v}),\quad {\hat H}_{gg}^0(v)\equiv e^{-v}P_{gg}^0(e^{-v}),\label{HLO}\\
{\hat H}_{qq}^1(v)&\equiv&e^{-v}P_{qq}^1(e^{-v}),\quad {\hat H}_{qg}^1(v)\equiv e^{-v}P_{qg}^1(e^{-v}),\quad {\hat H}_{gq}^1(v)\equiv e^{-v}P_{gq}^1(e^{-v}),\quad {\hat H}_{gg}^1(v)\equiv e^{-v}P_{gg}^1(e^{-v}).\label{HNLO}
\ea

The splitting functions $P_{qq}(e^{-v})$ and $P_{gg}(e^{-v})$ in the  integrals \eq{dFdtauoftau} and \eq{dGdtauoftau} involve the distribution $1/(1-e^{-v})_+$.  The integrals involving this term can be transformed to expressions that involve derivatives of ${\hat F}_s$ or ${\hat G}$ without the appearance of the singular factor $1/(1-e^{-v})$, for example, to  integrals of the form $\int_0^v\partial{\hat F}_s(w,\tau)/\partial{w} \ln[1-e^{-(v-w)}]dw$ and $\int_0^v\partial{\hat G}(w,\tau)/\partial{w} \ln[1-e^{-(v-w)}]dw$.  After this change, all of the integrals in Eqs.\ (\ref{dFdtauoftau}) and (\ref{dGdtauoftau}) are normal convolutions. 
By making a  Laplace transform in $v$, we can  factor these integrals, since the Laplace transform of a convolution is the product of the Laplace transform of the factors, i.e.,
\ba
{\cal L}\left[\int_0^v {\hat F}[w]{\hat H}[v-w]\,dw;s   \right]&=&{\cal L} [{\hat F}[v];s]\times {\cal L} [{\hat H}[v];s]\label{convolution}.
\ea

Defining the Laplace transforms 
\ba
f(s,\tau)&\equiv &{\cal L}\left[ \hat F_s(v,\tau);s\right],\qquad
g(s,\tau)\equiv {\cal L}[\hat G(v,\tau);s]
\ea
and noting that 
\ba
{\cal L}\left[{\partial \hat F_s \over\partial w}(w,\tau);s\right]=s f(s,\tau),\qquad
{\cal L}\left[{\partial \hat G \over\partial w}(w,\tau);s\right]=s g(s,\tau),
\ea
we can  factor the Laplace transforms of \eq{dFdtauoftau} and \eq{dGdtauoftau} into two coupled ordinary first order differential equations in the variable $\tau$ in  Laplace space $s$ with $\tau$-dependent coefficients. These can be written as
\ba
{\partial f\over \partial \tau }(s,\tau) &=&\left( \Phi_f^{LO}(s)+\frac{\alpha_s(\tau
)}{4\pi} \Phi_f^{NLO}(s)\right)f(s,\tau)+\left(\Theta_f^{LO}(s)+\frac{\alpha_s(\tau)}{4\pi}\Theta_f^{NLO}(s)\right)g(s,\tau)\label{df},\\
{\partial g\over \partial \tau }(s,\tau) &=&\left( \Phi_
g^{LO}(s)+\frac{\alpha_s(\tau)}{4\pi} \Phi_g^{NLO}(s)\right)g(s,\tau)+\left(\Theta_g^{LO}(s)+\frac{\alpha_s(\tau)}{4\pi}\Theta_g^{NLO}(s)\right)f(s,\tau),\label{dg}
\ea 
where we recall that the  the $Q^2$ dependence is through the function $\tau$, i.e., 
$
\tau(Q^2,Q_0^2)={1\over 4 \pi}\int_{Q_0^2}^{Q^2} \alpha_s(Q'^2)\,d\,\ln Q'^2.
$

The LO coefficients $\Phi^{LO}$ and $\Theta^{LO}$ are given by \cite{bdhm1}
\ba
\Phi^{LO}_f(s)&=&4 -{8\over 3}\left({1\over s+1}+{1\over s+2}+2\left(\psi(s+1)+\gamma_E\right)\right),\label{Phif}\\
\Theta_f^{LO}(s)&=&2n_f\left({1\over s+1}-{2\over s+2}+{2\over s+3} \right),\label{Thetaf}\\
\Phi_g^{LO}(s)&=&{33-2n_f \over 3} +12\left({1\over s}-{ 2\over s+1}+{1\over s+2}-{1 \over s+3}-\psi(s+1)-\gamma_E\right),\label{Phig}\\
\Theta_g^{LO}(s)&=&{8\over 3}\left({2\over s}-{2\over s+1}+{1\over s+2}.     \right),\label{Thetag}
\ea
Here $\psi(x)$ in Eqs.\ (\ref{Phif}) and (\ref{Phig}) is the digamma function and $\gamma_E=0.5772156\ldots$ is Euler's constant,  quantities that are introduced in the Laplace transform of the LO terms involving the distribution $1/(1-e^{-v})_+$ discussed above. The evaluation of the NLO  coefficients is straightforward, but too lengthy to be included in this note, and will be given, in the future, when we make numerical evaluations of $F_s(x,Q^2)$ and $G(x,Q^2)$ in NLO. 

In the case of LO, the $\tau$ dependence of the equations is trivial, and the equations can be solved  simply  \cite{bdhm1}, as already noted. The extra explicit dependence of the NLO terms on the right-hands of these equations on $\tau$ prevents a similar construction here. In order to  {\em decouple} and  {\em solve} \eq{df} and \eq{dg},  we  Laplace transform them a second time---this time with respect to the variable $\tau$---into $U$ space, i.e., we let 
\ba
{\cal F}(s,U)&\equiv &{\cal L}\left[f(s,\tau);U\right ],\qquad {\cal G}(s,U)\equiv {\cal L}\left[g(s,\tau);U\right ],\label{calFandG}\\
{\cal L}\left[{\partial f\over \partial \tau }(s,\tau);U\right]&=&U{\cal F}(s,U)-f_0(s),\qquad {\cal L}\left[{\partial g\over \partial \tau }(s,\tau);U\right]=U{\cal G}(s,U)-g_0(s),\label{dcalFanddcalG}
\ea
where now $s$ is simply  a parameter  in $U$ space. 

In $U$ space, we now write  the final desired coupled {\em algebraic} equations for ${\cal F}(s,U)$ and ${\cal G}(s,U)$ as 
\ba
U{\cal F}(s,U)-f_0(s)&=& \Phi_f^{LO}(s){\cal F}(s,U)+\Phi_f^{NLO}(s){\cal L}[\frac{\alpha_s(\tau
)}{4\pi} f(s,\tau);U]\nonumber\\
&&+\Theta_f^{LO}(s){\cal G}(s,U)+\Theta_f^{NLO}(s){\cal L}[\frac{\alpha_s(\tau)}{4\pi}g(s,\tau);U],\label{calFeqn}\\
U{\cal G}(s,U)-g_0(s)&=& \Phi_g^{LO}(s){\cal G}(s,U)+\Phi_g^{NLO}(s){\cal L}[\frac{\alpha_s(\tau
)}{4\pi} g(s,\tau);U]\nonumber\\
&&+\Theta_g^{LO}(s){\cal F}(s,U)+\Theta_g^{NLO}(s){\cal L}[\frac{\alpha_s(\tau)}{4\pi}f(s,\tau);U].\label{calGeqn}
\ea

For brevity, we replace the NLO ${\alpha_s(\tau)\over 4\pi}$ by $a(\tau)$ in \eq{calFeqn} and \eq{calGeqn}.
We can numerically show that an excellent approximation  to  $a(\tau)\equiv {\alpha_s(\tau)\over 4\pi}
$, accurate to a few parts in $10^4$, is given by the expression
\ba
a(\tau)&\approx a_0+ a_1e^{-b_1\tau},
\label{aoftau}
\ea
where the constants $a_0,a_1, b_1$ are found by a least squares fit to $a(\tau)$. We note in passing that this approximation is inspired by the fact that in LO, $\alpha_{s,LO}(\tau)$ is {\em exactly} given by the form $\alpha_{s,LO}(Q_0^2)e^{-b \tau}$. 

Using the value of $a(\tau)$ given by \eq{aoftau}, we can write the Laplace transforms ${\cal L}[\frac{\alpha_s(\tau)}{4\pi} f(s,\tau);U]$ and ${\cal L}[\frac{\alpha_s(\tau)}{4\pi} g(s,\tau);U]$ needed in \eq{calFeqn} and \eq{calGeqn} as
\ba
{\cal L}[\frac{\alpha_s(\tau)}{4\pi} f(s,\tau);U]&=&\sum_{j=0}^1a_j{\cal F}(s,U+b_j),\qquad
{\cal L}[\frac{\alpha_s(\tau)}{4\pi} g(s,\tau);U]=\sum_{j=0}^1 a_j{\cal G}(s,U+b_j),\label{transformoffalpha}
\ea
where $b_0=0$ is understood in \eq{aoftau} and \eq{transformoffalpha}.
 
After  introducing the simplifying notation
\ba
\Phi_f(s)\equiv \Phi_f^{LO}(s)+a_0\Phi_f^{NLO}(s),\qquad \Phi_g(s)\equiv \Phi_g^{LO}(s)+a_0\Phi_g^{NLO}(s),\label{Phis}\\
\Theta_f(s)\equiv \Theta_f^{LO}(s)+a_0\Theta_f^{NLO}(s),\qquad \Theta_g(s)\equiv \Theta_g^{LO}(s)+a_0\Theta_g^{NLO}(s),\label{Thetas}
\ea 
we can finally rewrite \eq{calFeqn} and \eq{calGeqn} as
\ba
\left[ U -\Phi_f(s)\right]{\cal F}(s,U)- \Theta_f(s){\cal G}(s,U)&=&f_0(s)+a_1\left[\Phi_f^{NLO}(s){\cal F}(s,U+b_1)+\Theta_f^{NLO}(s) {\cal G}(s,U+b_1)\right]
,\label{Feqn}\\
-\Theta_g(s){\cal F}(s,U)+\left[ U -\Phi_g(s)\right]{\cal G}(s,U)&=&g_0(s)+a_1\left[\Theta_g^{NLO}(s){\cal F}(s,U+b_1)+\Phi_g^{NLO}(s){\cal G}(s,U+b_1)\right]
,\label{Geqn}
\ea 
which we will solve iteratively, using $a_1$ as an expansion parameter. 

We  note that the $\Phi$'s and $\Theta$'s, as defined above, contain both LO {\em and} NLO terms.  
We further  point out that \eq{Feqn} and \eq{Geqn} are completely symmetric under the simultaneous transformations $f\leftrightarrow g$ and $\cal F \leftrightarrow \cal G$. We finally remark that $a_1$, the NLO  expansion parameter in our iterative solution  of Eqs. (\ref {Phis}) and (\ref {Thetas}), is quite small: $a_1=0.025, \ b_1=10.7$ for $M_c^2<Q^2\le M_b^2$ GeV$^2$ and $a_1=0.017, \ b_1=8.63$ for  $M_b^2<Q^2\le 10^5$ GeV$^2$, with the $a_0$ terms in \eq{Phis} and \eq{Thetas} being positive and about an order of magnitude smaller than the $a_1$ terms. 

We next consider the simple solutions to \eq{Feqn} and \eq{Geqn}, called  ${\cal F}_1(s,U)$ and ${\cal G}_1(s,U)$, that result from setting $a_1=0$, i.e., the equations
\ba
\left[ U -\Phi_f(s)\right]{\cal F}_1(s,U)- \Theta_f(s){\cal G}_1(U)&=&f_0(s),\label{F1eqn}\\
-\Theta_g(s){\cal F}_1(U)+\left[ U -\Phi_g(s)\right]{\cal G}_1(s,U)&=&g_0(s),\label{G1eqn}
\ea 
whose solutions are
\ba
{\cal F}_1(s,U)&=& \left[U-\Phi_g(s)\right]f_0(s)/D(U,s)+\Theta_f(s)g_0(s)/D(U,s) ,\label{F1}\\
{\cal G}_1(s,U)&=& \left[U-\Phi_f(s)\right]g_0(s)/D(U,s)+\Theta_g(s)f_0(s)/D(U,s) .\label{G1}
\ea

The denominator $D(U,s)$ in Eqs.\ (\ref{F1}) and (\ref{G1}) is just the determinant of the coefficients of  ${\cal F}(s,U)$ and ${\cal G}(s,U)$ in Eqs. (\ref{Feqn}) and (\ref{Geqn}),  
\ba
D(U,s) &=& \Phi_f(s)\Phi_g(s)-\Theta_f(s)\Theta_g(s)-\left[ \Phi_f(s)+\Phi_g(s)\right] U+U^2 \nonumber \\
\label{D(U,s)}
&=& \left(U-\frac{1}{2}\left(\Phi_f(s)+\Phi_g(s)\right)-\frac{1}{2}R(s)\right)\left(U-\frac{1}{2}\left(\Phi_f(s)+\Phi_g(s)\right)+\frac{1}{2}R(s)\right),
\ea
where $R(s) \equiv \sqrt{\left(\Phi_f(s)-\Phi_g(s)\right)^2+4 \Theta_f(s)\Theta_g(s)}$. 
The zeros of $D(U,s)$ lead to simple poles in ${\cal F}_1$ and ${\cal G}_1$ in the $U$ plane.  These functions have no other singularities,  and decrease as $1/|U|$ for $|U|\rightarrow\infty$.  The inverse Laplace transforms of ${\cal F}_1(s,U)$ and ${\cal G}_1(s,U)$, denoted by $f_1(s,\tau)$ and $g_1(s,\tau)$, are therefore well defined and simple to calculate. We will write them as
\ba
f_1(s,\tau)&= &k_{ff_1}(s,\tau)f_0(s)+k_{fg_1}(s,\tau) g_0(s),\qquad
g_1(s,\tau)= k_{gg_1}(s,\tau)g_0(s)+k_{gf_1}(s,\tau) f_0(s)\label{f1andg1},
\ea 
where the coefficient functions in the solution are
\ba
k_{ff_1}(s,\tau)&\equiv&e^{\frac{{\tau }}{2}\left(\Phi_f(s) +\Phi_g(s)\right)}\left[\cosh\left (  {\tau \over 2}R(s)\right) +\frac{\sinh\left({\tau\over2}R(s)\right)}{R(s)} \left(\Phi_f(s)-\Phi_g(s)\right)\right],\label{kff}\\
k_{fg_1}(s,\tau)&\equiv &e^{ {\tau\over 2}\left(\Phi_f(s)+\Phi_g(s)\right) }{2\sinh\left ( {\tau \over 2}R(s) \right )\over R(s)}\,\Theta_f(s),\label{kfg}\\
k_{gg_1}(s,\tau)&\equiv &e^{{\tau \over2}\left(\Phi_f(s) +\Phi_g(s)\right)}\left[\cosh\left (  {\tau \over 2}R(s)\right) -\frac{\sinh\left({\tau\over2}R(s)\right)}{R(s)} \left(\Phi_f(s)-\Phi_g(s)\right)\right],\label{kgg}\\
k_{gf_1}(s,\tau)&\equiv&e^{ {\tau\over 2}\left(\Phi_f(s)+\Phi_g(s)\right) }{2\sinh\left ( {\tau \over 2}R(s) \right )\over R(s)}\,\Theta_g(s).\label{kgf}
\ea
We comment  that this solution has small NLO terms in it, arising from the $a_0$ term in  \eq{Phis} and  \eq{Thetas}. However, it is {\em identical} in form to the LO solution we gave in Ref. \cite{bdhm1},  and reduces to it if we set $a_0=0$.  

We  next construct  an iterative solution to Eqs.\ (\ref{Feqn}) and (\ref{Geqn}) for ${\cal F}$ and ${\cal G}$. We start by substituting the known functions ${\cal F}_1$ and ${\cal G}_1$ for ${\cal F}$ and ${\cal G}$ on the right hand sides of the equations, and then re-solve the equations to obtain the next approximations ${\cal F}_2$ and ${\cal G}_2$ for ${\cal F}$ and ${\cal G}$, and then repeat the process. For  the first step, we make the replacements
\ba
{\cal F}(s,U+b_1)\rightarrow {\cal F}_1(s,U+b_1),\qquad {\cal G}(s,U+b_1)\rightarrow {\cal G}_1(s,U+b_1),\qquad \label{FtoF1GtoG1}
\ea
on the right-hand sides of Eqs.\ (\ref{Feqn}) and (\ref{Geqn}) to obtain
 our first iterative equations for ${\cal F}_2(s,U)$ and ${\cal G}_2(s,U)$, 
\ba
\left[ U -\Phi_f(s)\right]{\cal F}_2(s,U)- \Theta_f(s){\cal G}_2(s,U)&=&f_0(s) +a_1\left[\Phi_f^{NLO}(s){\cal F}_1(s,U+b_1)+\Theta_f^{NLO}(s) {\cal G}_1(s,U+b_1)\right],
\label{F2eqn1}\\
-\Theta_g(s){\cal F}_2(s,U)+\left[ U -\Phi_g(s)\right]{\cal G}_2(s,U)&=&g_0(s)+a_1\left[\Theta_g^{NLO}(s){\cal F}_1(s,U+b_1)+\Phi_g^{NLO}(s){\cal G}_1(s,U+b_1)\right].
\label{G2eqn1}
\ea 
The functions ${\cal F}_1(s,U)$ and ${\cal G}_1(s,U)$ are given analytically by \eq{F1} and \eq{G1}, respectively, so that we know them at the argument $(s,U+b_1)$, needed in the right hand sides of our iterative equations. 

Since the functions on the right-hand sides of Eqs.\ (\ref{F2eqn1}) and (\ref{G2eqn1}) are known, and the left-hand sides have the same structure as Eqs.\ (\ref{F1eqn}) and (\ref{G1eqn}), their solutions  can be obtained by the substitutions
\ba
f_0(s)&\rightarrow  &f_0(s) + a_1\left[\Phi_f^{NLO}(s){\cal F}_1(s,U+b_1)+\Theta_f^{NLO}(s) {\cal G}_1(s,U+b_1)\right],\label{fsubstitution}\\
g_0(s)&\rightarrow&g_0(s)+ a_1\left[\Theta_g^{NLO}(s){\cal F}_1(s,U+b_1)+\Phi_g^{NLO}(s){\cal G}_1(s,U+b_1)\right]\label{gsubstitution},
\ea
on the right hand sides of Eqs.\  (\ref{F1}) and (\ref{G1}). The leading $f_0(s)$ and $g_0(s)$ reproduce ${\cal F}_1(s,U)$ and ${\cal G}_1(s,U)$. The added terms, proportional to the expansion parameter $a_1$, are more complicated expressions that are  rational functions in $U$, whose numerators are a second-order polynomial and whose denominators are the factorable product  $D(U,s)D(U+b_1,s)$. The functions ${\cal F}_2$ and ${\cal G}_2$ therefore have an extra pole in $U$ that is displaced along the real axis from the poles of ${\cal F}_1$ and ${\cal G}_1$ by the amount $b_1$.  Since this is the only new singularity, whose terms decrease at least as rapidly as $1/U^2$ for $U\rightarrow\infty$,  the overall behavior of the  iterated solution decreases at least as rapidly as $1/|U|$, so that the inverse Laplace transforms needed can be calculated analytically.

We can again write the  results for the inverse transforms $f_2(s,\tau)$ and $g_2(s,\tau)$ in terms of the initial distributions $f_0(s)$ and $g_0(s)$ as in \eq{f1andg1}, but with the coefficient functions $k$ now sums of the original expressions in Eqs. (\ref{kff})-(\ref{kgf}) and terms that depend linearly on the coefficient $a_1$ in \eq{aoftau} as well as on $s$ and $\tau$.   The coefficient $a_0$ has been incorporated into the definitions of the $\Phi$'s and $\Theta$'s in Eqs.\ (\ref{Phis}) and (\ref{Thetas}), so it does not appear explicitly. 

Continuing, we find the $k^{\rm th}$ iterated solution to Eqs.\ (\ref{Feqn}) and (\ref{Geqn}) for ${\cal F}(s,U)$  and ${\cal G}(s,U)$ by making the substitutions 
\ba
f_0(s)&\rightarrow &f_0(s)+a_1\left[\Phi_f^{NLO}(s){\cal F}_k(s,U+b_1)+\Theta_f^{NLO}(s){\cal G}_k(s,U+b_1)\right], \\
g_0(s)&\rightarrow &f_0(s)+a_1\left[\Phi_g^{NLO}(s){\cal G}_k(s,U+b_1)+\Theta_g^{NLO}(s){\cal F}_k(s,U+b_1)\right]\label{g0new}
\ea
in the right-hand sides of Eqs. ({\ref{F1})  and (\ref {G1}) and replacing  ${\cal F}_{1}(s,U)$ and ${\cal G}_{1}(s,U)$ on the left-hand side by   ${\cal F}_{k+1}(s,U)$ and ${\cal G}_{k+1}(s,U)$. The resulting expressions for ${\cal F}_{k+1}$ and ${\cal G}_{k+1}$   add new terms proportional to $a_1^k$, which again are rational functions of $U$, with denominator  of higher power in $U$ than the numerator. All the terms in
 ${\cal F}_{k+1}(s,U),\ {\cal G}_{k+1}(s,U)$ decrease at least as rapidly as $1/|U|$ for $|U|\rightarrow\infty$, and the only singularities are poles at known locations, so  we can again calculate the Laplace  inversion  from $U$ space to $\tau$ space analytically. At each stage, 
we can  write the inverse transforms as
\ba
f(s,\tau)&=&k_{ff}(a_1,b_1,s,\tau)f_0(s)+k_{fg}(a_1,b_1,s,\tau)g_0(s),\\
 g(s,\tau)&=&k_{gg}(a_1,b_1,s,\tau)g_0(s)+k_{gf}(a_1,b_1,s,\tau)f_0(s),\label{fandg}
\ea
with the functions $k(a_1,b_1,s,\tau)$  expressed as  power series in the NLO expansion parameter $a_1$ whose coefficients are analytic functions of $s$ and $\tau$. These expressions rapidly become too complicated and too lengthy to reproduce here, but are easily calculated using a program such as Mathematica \cite{Mathematica}.   

After numerical Laplace inversion \cite{inverseLaplace1,inverseLaplace2} of the $k$'s from $s$ to $v$ space, suppressing their explicit dependence on $a_1$ and $b_1$, we define their Laplace inverses as
\ba
K_{FF}(v,\tau)&\equiv & {\cal L}^{-1}[k_{ff}(s,\tau);v],\qquad K_{FG}(v,\tau)\equiv  {\cal L}^{-1}[k_{fg}(s,\tau);v],\label{KFF&KFG}\\
K_{GG}(v,\tau)&\equiv & {\cal L}^{-1}[k_{gg}(s,\tau);v],\qquad K_{GF}(v,\tau)\equiv  {\cal L}^{-1}[k_{gf}(s,\tau);v],\label{KGG&KGF}
\ea
so that we can write the {\em decoupled solutions} in $(v,\tau)$ space as the convolutions
\ba
{\hat F}_s(v,Q^2)&\equiv&\int_0^v K_{FF}(v-w,\tau(Q^2,Q_0^2)){\hat F}_{s0}(w)\,dw+\int_0^v K_{FG}(v-w,\tau(Q^2,Q_0^2)){\hat G}_{0}(w)\,dw, \label{Fhatofvtau}\\
{\hat G}(v,Q^2)&\equiv&\int_0^v K_{GG}(v-w,\tau(Q^2,Q_0^2)){\hat G}_{0}(w)\,dw+\int_0^v K_{GF}(v-w,\tau(Q^2,Q_0^2)){\hat F}_{s0}(w)\,dw. \label{Ghatofvtau}
\ea 
Finally, recalling that $v\equiv \ln(1/x)$, we can transform  the above solutions back into the usual space, Bjorken-$x$ and virtuality $Q^2$, enabling us to write the NLO {\em decoupled}  solutions, $F_s(x,Q^2)$ and $G(x,Q^2)$, which require only a knowledge of $F_{s0}(x)$ and $G(x)$ at $Q_0^2$, where evolution is started.

In order to insure continuity across the boundaries $Q^2=M_c^2$ and $M_b^2$, we  first evolve from $Q_0^2$ (where, e.g.,  $Q_0^2=1$ GeV$^2$ for the MSTW group \cite{MSTW1}) to $M_c^2$ and use our evolved values of ${\hat F}_{s0}(v)$ and  ${\hat G}_{0}(v)$ for a {\em new} starting values of ${\hat F}_{s0}(v)$ and  ${\hat G}_{0}(v)$. We  then evolve to $M_b^2$, repeating the process, thus insuring continuity of $F_{s}(x,Q^2)$ and $G_{s}(x,Q^2)$ at the boundaries where $n_f$ changes.


\section{Non-singlet sector}
For {\em non-singlet} distributions $F_{ns}(x,Q^2)$, such as for valence quarks, $D_{\rm val}=x\left(d(x,Q^2)-{\bar d}(x,Q^2)\right)$---the {\em difference}  between quark distributions---we can schematically write the logarithmic derivative of $F_{ns}$ as the convolution of $F_{ns}(x,Q^2)$ with the non-singlet splitting functions, $P_{qq}^{LO,ns}(x)$ and $P_{qq}^{NLO,ns}(x)$, for LO and NLO, respectively,   (using the convolution symbol $\otimes$), i.e., 
\ba
{4\pi\over \alpha_s(Q^2)}{\partial F_{ns}\over \partial \ln (Q^2)}(x,Q^2)&=&F_{ns}\otimes\left[ P_{qq}^{LO,ns}+{\alpha_s(\tau)\over 4\pi}P_{qq}^{NLO,ns}\right](x,Q^2).\label{nonsingletinQsq_1}
\ea

After changing to the variable $v=\ln(1/x)$ and the variable $\tau$,
we write
\ba
{\partial {\hat F_{ns}}\over \partial \tau}(v,\tau)&=&\int_0^v{\hat F}_{ns}(w,\tau) e^{-(v-w)}\left[ P_{qq}^{LO,ns}(v-w)+{\alpha_s(\tau)\over 4\pi}P_{qq}^{NLO,ns}(v-w)\right]d\,w.\label{nonsingletinQsq_2}
\ea
The comments that we made in Sec.\ II about  integrals that involve the distribution $1/(1-e^{-v})_+$ also apply here. 

Going to Laplace space $s$, we obtain a linear differential equation in $\tau$ for the transform $f_{ns}(s,\tau)$. This has the simple solution
\ba
f_{ns}(s,\tau)&=&e^{\tau \Phi_{ns}(s)}f_{ns0}(s),\qquad \Phi_{ns}(s)\equiv \Phi_{ns}^{LO}(s)+{\tau_2\over \tau}\Phi_{ns}^{NLO}(s),\label{solutionandPhinsofs}
\ea
where
\ba 
\tau_2&\equiv& {1\over 4\pi}\int_0^\tau\alpha_s(\tau')\,d\tau' ={1\over (4\pi)^2}\int_{Q^2_0}^{Q^2}\alpha_s^2(Q'^2)\, d \ln Q'^2,\label{tau2ofQsq}
\ea
and
\ba
\Phi_{ns}^{LO}(s)&\equiv&{\cal L}\left [e^{-v}P_{qq}^{LO,ns}(e^{-v});s\right ],\qquad\Phi_{ns}^{NLO}(s)\equiv{\cal L}\left [e^{-v}P_{qq}^{NLO,ns}(e^{-v});s\right ].\label{PhiNs0and1}
\ea
We note that in LO, $\Phi_{ns}(s)=\Phi_f^{LO}(s)$, where  $\Phi_f^{LO}(s)$ has been written out explicitly in \eq{Phif}. Again, the evaluation of $\Phi_{ns}^{NLO}(s)$ is straightforward, but too lengthy to be shown here.  

We can find {\em any} non-singlet solution, $F_{ns}(x,Q^2)$, by  using the non-singlet kernel $K_{ns}(v)\equiv {\cal L}^{-1}\left[e^{\tau \Phi_{ns}(s)};v   \right]$ in the Laplace  convolution relation  
\ba
{\hat F}_{ns}(v,\tau)=\int _0^v K_{ns}(v-w,\tau)\hat F_{ns0}(w)\,dw\label{Fnsofv},
\ea
and then returning to $(x,Q^2)$ space. 

In order to insure the continuity of $F_{ns}(x,Q^2)$ where $n_f$ changes, we renormalize the starting values ${\hat F}_{ns0}(v)$ at the boundaries $M_c^2$ and $M_b^2$, as described previously in a similar context for singlet distributions. 

\subsection{Comparison of non-singlet theory with NLO MSTW non-singlet valence quark  distributions}
As an example of the application of this technique, we will compare two  $x$-space non-singlet valence quark distribution functions $F_{ns}(x,Q^2)$ calculated from  \eq{Fnsofv} with the published MSTW values \cite{MSTW1}. In \eq{Fnsofv}, we use $Q_0^2=1 $ GeV$^2$, the MSTW starting value for evolution, to  construct  ${\hat F}_{ns0}(v)$ from  the published NLO MSTW \cite{MSTW1} quark distributions.  We use the MSTW values  $M_c=1.40$ GeV, $M_b=4.75$ GeV,  together with  the MSTW NLO definition of $\alpha_s(Q^2)$, adjusted to be continuous at the boundaries $Q^2=M_b^2$ and $M_c^2$,  with $\alpha_s(1\ {\rm GeV}^2)=0.49128$  and  $\alpha_s(M_Z^2)=0.12018$ \cite{MSTW1}.

\subsubsection{The NLO non-singlet $d$ quark valence distribution $D_{\rm val}=x\left(d(x,Q^2)-{\bar d}(x,Q^2)\right)$\label{Dval}}
In Fig. \ref{fig:dvalence}, we show the results obtained by evolving the non-singlet $d$ quark valence distribution, $D_{\rm val}=x\left(d(x,Q^2)-{\bar d}(x,Q^2)\right)$,  from $Q_0^2=1$ GeV$^2$, for $Q^2 = 5\ ,20,\ 100$ and $M_z^2$ GeV$^2$.
 The published MSTW \cite{MSTW1} curves are: $Q^2 = 5$ GeV$^2$, solid blue; $Q^2 = 20$ GeV$^2$, dashed green; $Q^2 = 100$ GeV$^2$, dot dashed red; $Q^2 = M_z^2$ GeV$^2$, large dashed black.  The dots are our evolution results for NLO non-singlet $D_{\rm val}=x\left(d(x,Q^2)-{\bar d}(x,Q^2)\right)$ from \eq{Fnsofv} (converted to $x$-space), using the NLO MSTW values for $F_{ns0}(x)$, where $Q^2_0=1$ GeV$^2$. Let us define the fractional error, ${\rm frac. \ err.}\equiv 1-D_{\rm val}({\rm calculated})/D_{\rm val}({\rm MSTW})$,  at $x=0.135$, a point near the peaks of the curves in Fig. \ref{fig:dvalence}. The reproduction of the published MSTW data is excellent. We find that at $Q^2= 5 $ GeV$^2$, frac. err. = -0.004 and at $Q^2= M_z^2$, frac. err. = +0.004. 
\begin{figure}[h]
\begin{center}
\mbox{\epsfig{file=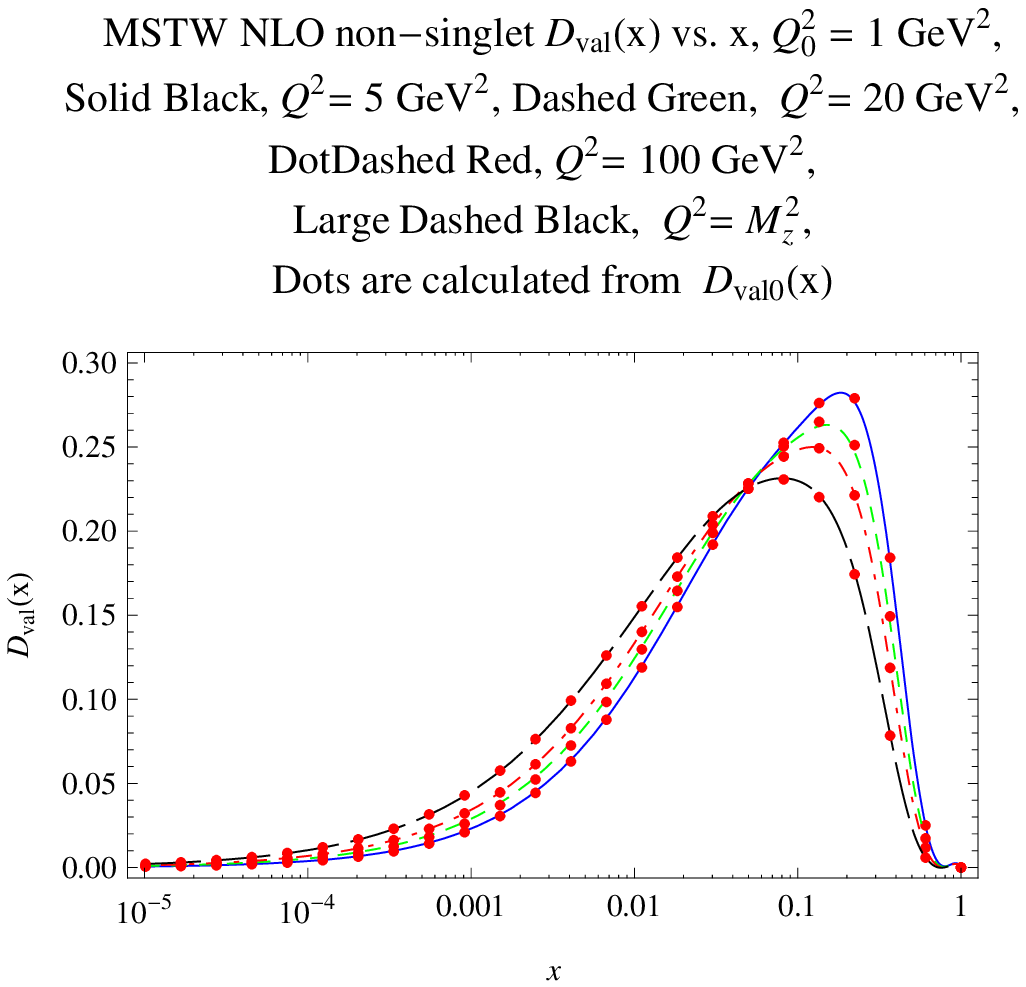
,width=4.8in%
,bbllx=0pt,bblly=0pt,bburx=317pt,bbury=185pt,clip=%
}}
\end{center}
\caption[]{
The NLO MSTW \cite{MSTW1} non-singlet valence distribution, $D_{\rm val}=x\left(d(x,Q^2)-{\bar d}(x,Q^2)\right)$, for $Q^2 = 5\ ,20,\ 100$ and $M_z^2$ GeV$^2$. The published MSTW \cite{MSTW1} curves are: $Q^2 = 5$ GeV$^2$, solid blue; $Q^2 = 20$ GeV$^2$, dashed green; $Q^2 = 100$ GeV$^2$, dot dashed red; $Q^2 = M_z^2$ GeV$^2$, large dashed black.  The dots are the evolution results for NLO non-singlet $D_{\rm val}=x\left(d(x,Q^2)-{\bar d}(x,Q^2)\right)$ from \eq{Fnsofv} (converted to $x$-space), using the NLO MSTW values for $F_{ns0}(x)$, where $Q^2_0=1$ GeV$^2$. 
} 
\label{fig:dvalence}
\end{figure}
\nopagebreak
\begin{figure}[h]
\begin{center}
\mbox{\epsfig{file=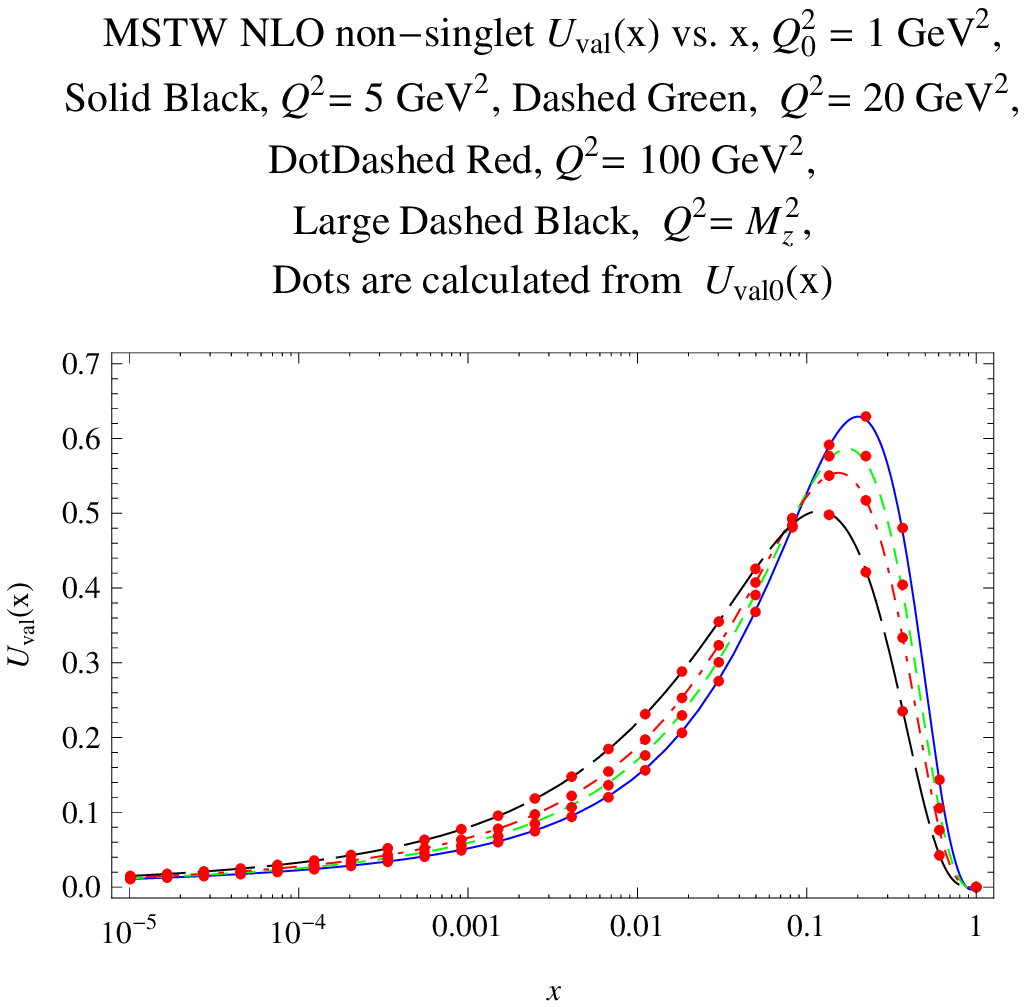
,width=4.8in%
,bbllx=0pt,bblly=0pt,bburx=317pt,bbury=190pt,clip=%
}}
\end{center}
\caption[]{
The NLO MSTW \cite{MSTW1} non-singlet valence distribution, $U_{\rm val}=x\left(u(x,Q^2)-{\bar u}(x,Q^2)\right)$, for $Q^2 = 5\ ,20,\ 100$ and $M_z^2$ GeV$^2$. The published MSTW \cite{MSTW1} curves are: $Q^2 = 5$ GeV$^2$, solid blue; $Q^2 = 20$ GeV$^2$, dashed green; $Q^2 = 100$ GeV$^2$, dot dashed red; $Q^2 = M_z^2$ GeV$^2$, large dashed black.  The dots are the evolution results for NLO non-singlet valence distribution,  $U_{\rm val}=x\left(u(x,Q^2)-{\bar u}(x,Q^2)\right)$ from \eq{Fnsofv} (converted to $x$-space), using the NLO MSTW values for $F_{ns0}(x)$, where $Q^2_0=1$ GeV$^2$. 
} 
\label{fig:Uvalence}
\end{figure}
\subsubsection{The NLO non-singlet $u$ quark valence distribution $U_{\rm val}=x\left(u(x,Q^2)-{\bar u}(x,Q^2)\right)$}

In Fig. \ref{fig:Uvalence}, we show the results obtained by evolving the non-singlet distribution valence distribution for the $u$ quark,  $U_{\rm val}(x,Q^2)=x\left(u(x,Q^2)-{\bar u}(x,Q^2)\right)$ from $Q_0^2=1$ GeV$^2$, for $Q^2 = 5\ ,20,\ 100$ and $M_z^2$ GeV$^2$.
 The published MSTW \cite{MSTW1} curves are: $Q^2 = 5$ GeV$^2$, solid blue; $Q^2 = 20$ GeV$^2$, dashed green; $Q^2 = 100$ GeV$^2$, dot dashed red; $Q^2 = M_z^2$ GeV$^2$, large dashed black.  The dots are our evolution results for NLO non-singlet $u$ quark valence distribution $U_{\rm val}(x,Q^2)=x\left(u(x,Q^2)-{\bar u}(x,Q^2)\right)$ from \eq{Fnsofv} (converted to $x$-space), using the NLO MSTW values for $F_{ns0}(x)$, where $Q^2_0=1$ GeV$^2$. Again, the reproduction of the published MSTW data is excellent. The fractional errors at $x=0.135$, defined in Section \ref{Dval},  are:  frac. err. = -0.003 at $Q^2= 5 $ GeV$^2$ and  frac. err. = +0.004 at $Q^2= M_z^2$ GeV$^2$.

\section{Conclusions}
For the singlet sector of pQCD, we have solved the coupled NLO DGLAP equations and found  NLO {\em decoupled} analytic solutions for $F_s(x,Q^2)$ and $G(x,Q^2)$, an extension of our earlier work for LO \cite{bdhm1}. All that is required is knowledge of the initial distributions $F_{s0}(x)$ and G$_0(x)$, at $Q^2=Q_0^2$, where $Q_0^2$ is the starting value for the evolution.  For the non-singlet sector, we have successfully solved the NLO evolution equation for $F_{ns}(x,Q^2)$, again in terms of $F_{ns0}(x)$, the value of $F_{ns}(x,Q^2)$ at $Q_0^2$. We illustrated this numerically for NLO, calculating the non-singlet valence quark distributions $U_{\rm val}=x\left(u(x,Q^2)-{\bar u}(x,Q^2)\right)$ and $D_{\rm val}=x\left(d(x,Q^2)-{\bar d}(x,Q^2)\right)$ for a very large range of $x$ and $Q^2$, in excellent agreement with the  NLO published MSTW \cite{MSTW1} values. We note that these techniques can be extended to arbitrary order in the strong coupling constant $\alpha_s(Q^2)$, for both the singlet and non-singlet sector. 

The results presented here are basically analytic, thus {\em eliminating} the need for simultaneous numerical  solutions of  the singlet and non-singlet DGLAP equations on a two-dimensional lattice in $x$ and $Q^2$. They provide {\em new tools} for studying pQCD; for example, they can be used  to examine directly the sensitivity of an {\em  individual\ }  evolved distribution to the assumed shapes of its starting distribution. In the future,  we hope to apply these techniques to a global fit of experimental $F_2^{\gamma p}(x,Q^2)$ data to determine in LO, a gluon starting distribution, and in NLO, approximate $F_s$ and gluon starting distributions. Since these  starting distributions will be determined by experimental data, they will be free of predetermined shape hypotheses; this will allow us to find new gluon distributions that are critically needed for the interpretation of results from the Large Hadron Collider.
    
\section{Acknowledgments}
The authors would like to  thank the Aspen Center for Physics for its hospitality during the time parts of this work were done. P. Ha would like to thank Towson University Fisher College of Science and Mathematics for travel support.
   D.W.M. received travel support from DOE Grant No. DE-FG02-04AR41308.


\bibliography{gluonsPRD.bib}

\end{document}